\documentclass[preprint,12pt]{elsarticle}

\usepackage{amssymb}
\usepackage{graphicx}%
\usepackage{multirow}%
\usepackage{amsmath,amssymb,amsfonts}%
\usepackage{amsthm}%
\usepackage{mathrsfs}%
\usepackage[title]{appendix}%
\usepackage{xcolor}%
\usepackage{textcomp}%
\usepackage{manyfoot}%
\usepackage{booktabs}%
\usepackage{algorithm}%
\usepackage{algorithmicx}%
\usepackage{algpseudocode}%
\usepackage{listings}%
\usepackage[english]{babel}
\usepackage{graphicx}
\usepackage{amssymb}
\usepackage{hyperref}
\usepackage{graphicx}
\usepackage{subfigure,dcolumn}
\usepackage{graphicx,xcolor,float}
\usepackage[T2A,T1]{fontenc}
\usepackage[english]{babel}
\usepackage{hyperref}
\usepackage{float}
\usepackage{amssymb}
\usepackage{natbib}
\usepackage{float}
\usepackage{multirow}
\journal{NIM-A}

\begin{document}

\begin{frontmatter}



\title{The Pixel Charging-up effect in Gas Micro-Pixel Detectors}


\author[label1,label2]{Di-Fan Yi}
\affiliation[label1]{School of Physical Science, University of Chinese Academy of Sciences, Beijing, 100049, China}
\affiliation[label2]{School of Physical Science and Technology, Guangxi University, Nanning 530004, China}
\author[label1]{Qian Liu}
% \email[Corresponding author, ]{Qian Liu, liuqian@ucas.ac.cn}
\author[label2]{Hong-Bang Liu}
\author[label2]{Fei Xie}
\author[label2]{Huan-Bo Feng}
\author[label2]{Lin Zhou}
\author[label2]{Zu-ke Feng}
\author[label1]{Yang-Heng Zheng}

% \author{}

% \affiliation{organization={},%Department and Organization

\begin{abstract}
This study investigates the charging-up effect on the Topmetal-II$^-$ chip in Gas Micro-Pixel Detectors(GMPD). It is found that this effect differs from the charging-up typically observed in gas detector multiplier devices and increases the relative gain of the detector. The research indicates that this effect originates from the accumulation of charges on the insulating layer of the chip's pixel surface. Iterative simulations using COMSOL and GARFIELD++ are employed to model the variation of detector relative gain with the charging-up effect, and a simple yet effective model is proposed, which aligns well with experimental data. The feasibility of validating the deposition of resistive materials and adjusting the local voltage distribution on the chip to suppress charging-up effects and enhance the relative gain is also verified.
\end{abstract}


% \begin{highlights}
% \end{highlights}


\end{frontmatter}


\section{Introduction}
\label{}
Topmetal-II$^-$ is a charge-sensitive pixel chip widely used in particle track detection applications\cite{LI2021165430}. GMPD is a soft X-ray polarimeter developed by Guangxi University\cite{Feng_2023}, in which Topmetal-II$^-$ is utilized as the photoelectron track imaging component. GMPD will serve as a prototype for the space project POLAR-2/LPD\cite{inproceedingsAngelis,Fengzk_2024} payload, completing on-orbit performance verification and scientific observation tests. 

With the accumulation of working time, charging-up occurs on certain components of the detector, forming an electric field opposite to the drift or avalanche multiplication electric field, thereby affecting the instrument's performance. Charging-up effects have been extensively studied in charge multiplication devices such as gas electron multiplier (GEM)\cite{ALFONSI20126}, thick gas electron multipliers (THGEM)\cite{Song_2020}, gas microchannel plate (GMCP)\cite{FENG2023168499}, and MicroMigas\cite{CHEFDEVILLE2021165268}, which lead to a decrease in the electric field strength in the multiplication region, resulting in a reduction in detector gain. Numerous studies have indicated that the gain of gas detectors exhibits time evolution when bias is applied. Research has also shown that the gain of GEM increases over time\cite{azmoun2006study,hauer2020measurements,alfonsi2012simulation}, while the gain of THGEM initially decreases and then increases over time\cite{alexeev2015gain,correia2018simulation,pitt2018measurements}. The charge accumulation effect is commonly considered to be a significant factor influencing gain.

For GMPD, the charging-up effect occurs on the pixels of the Topmetal-II$^-$ chip. The charging-up effect on GMPD has many significant differences compared to the charging-up effects occurring on the aforementioned multiplication devices: the charging-up effect on Topmetal-II$^-$ increases the effective gain of the detector, and unless artificial measures are taken, the charge on Topmetal-II$^-$ will hardly dissipate on its own. This charging-up effect affects the energy resolution of the detector. More importantly, if the intensity of the incident source has a non-zero gradient, causing the charging-up effect to occur non-uniformly across the entire chip surface, it leads to inconsistent gains in different regions, ultimately affecting the reconstruction of the incident photon's polarization angle and resulting in significant residual modulation\cite{BALDINI2021102628,Rankin_2022}.

This paper investigates the charging-up mechanism of Topmetal-II$^-$ and proposes a simple and experimentally consistent charging-up model through comparisons and validations with experimental and simulated data. Section \ref{sec:structure} of this paper introduces the polarization detection principles of GMPD, the working principles, and performance indicators of Topmetal-II$^-$. Section \ref{sec:simulation} presents the algorithm for simulating charging-up in Topmetal-II$^-$. Section \ref{sec:modeling} compares experimental data with simulation results. Section \ref{sec:impact} discusses the impact of charging-up effects on residual modulation\cite{yi2024effectivenessstudycalibrationcorrection} in polarization detection. Section \ref{sec:suppressing} discusses the inhibitory effect of coating on charge accumulation. The final section \ref{sec:conclusion} provides a summary and outlook.

\section{Detector structure}
\label{sec:structure}
\begin{figure}[htbp]
  \centering
  \subfigure[]{\includegraphics[width=0.5\textwidth]{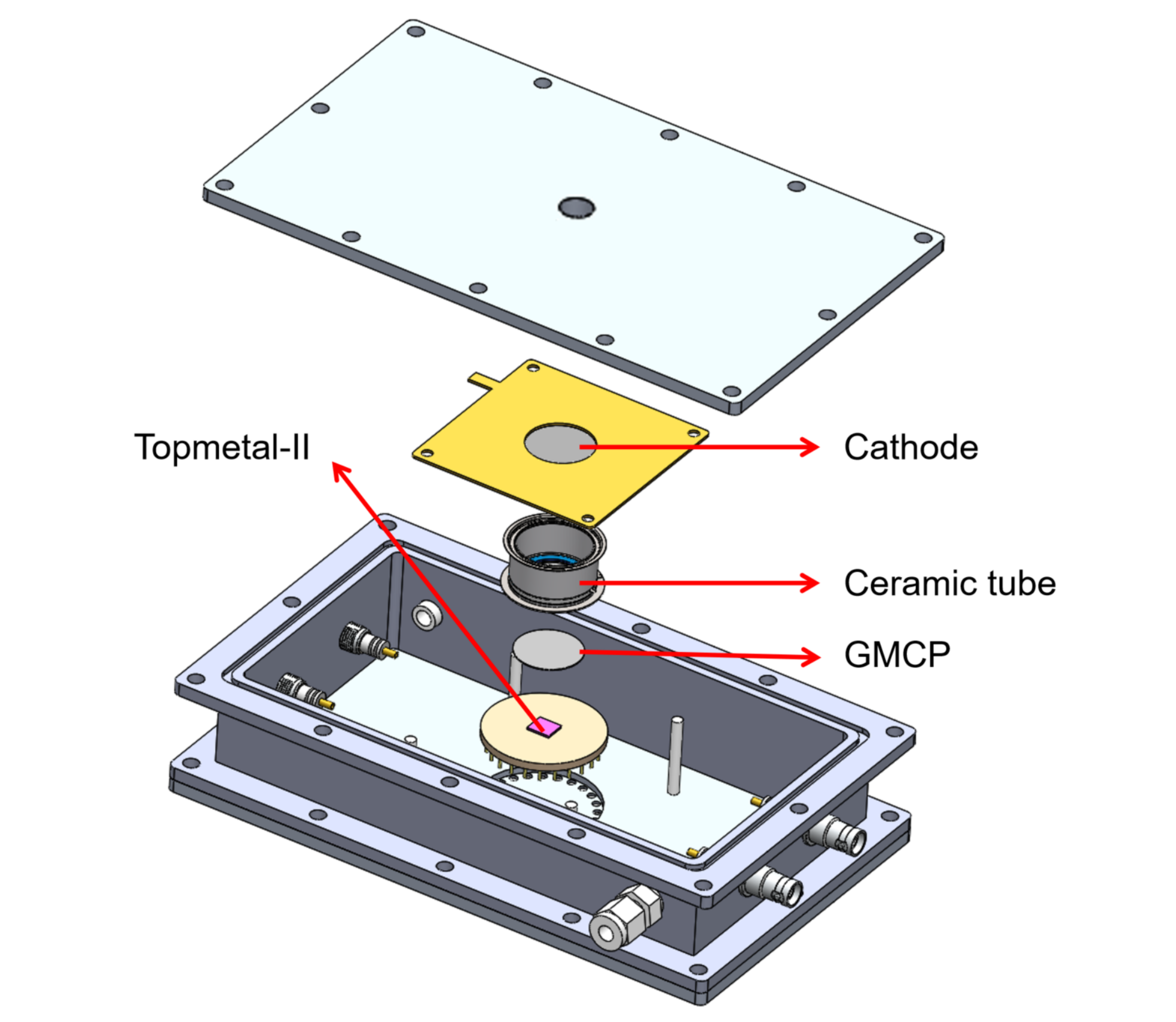}}
  \hfill
  \subfigure[]{\includegraphics[width=0.3\textwidth]{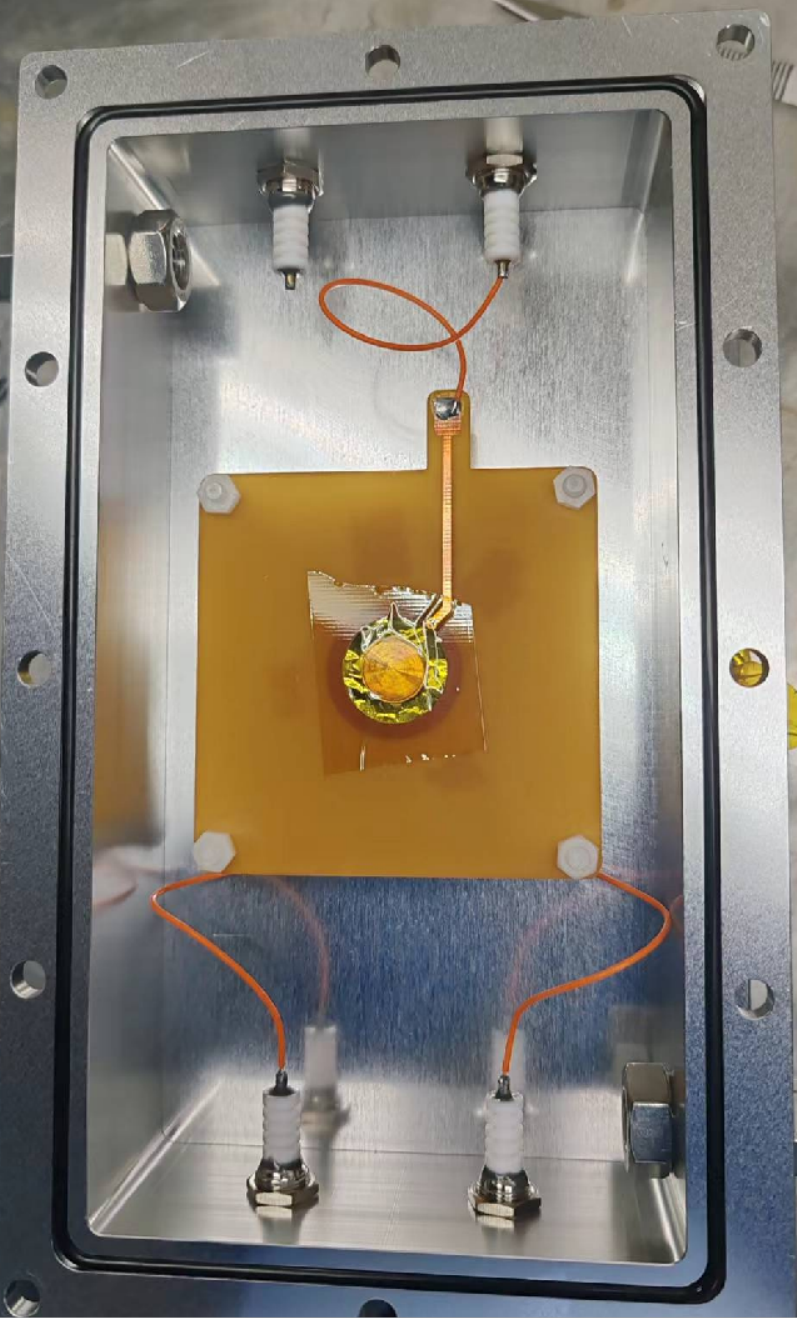}}
  \caption{(a) GMPD exploded view. (b) GMPD physical top view.}
  \label{fig:Dec_Struc}
\end{figure}

The prototype of the GMPD detector is shown in in Figure \ref{fig:Dec_Struc}. The entire detection unit is sealed in an aluminum flow gas chamber, with a constant pressure mixture of 30\% He and 70\% dimethyl ether (DME) flowing into the chamber at a rate of 30\,mL/min. The top layer of the detection unit is a titanium metal thin film cathode; in the middle is a drift region supported by a 1.4\,cm ceramic cylinder structure, with a 400\,$\mu$m thick GMCP fixed at the bottom of the ceramic cylinder, serving as the electron multiplication region. The Topmetal-II$^-$ chip is fixed on the base of the chamber, with a height of 3.4\,mm above the chip for the induction region.

\begin{figure}[htbp]
\includegraphics
  [width=1.\hsize]
  {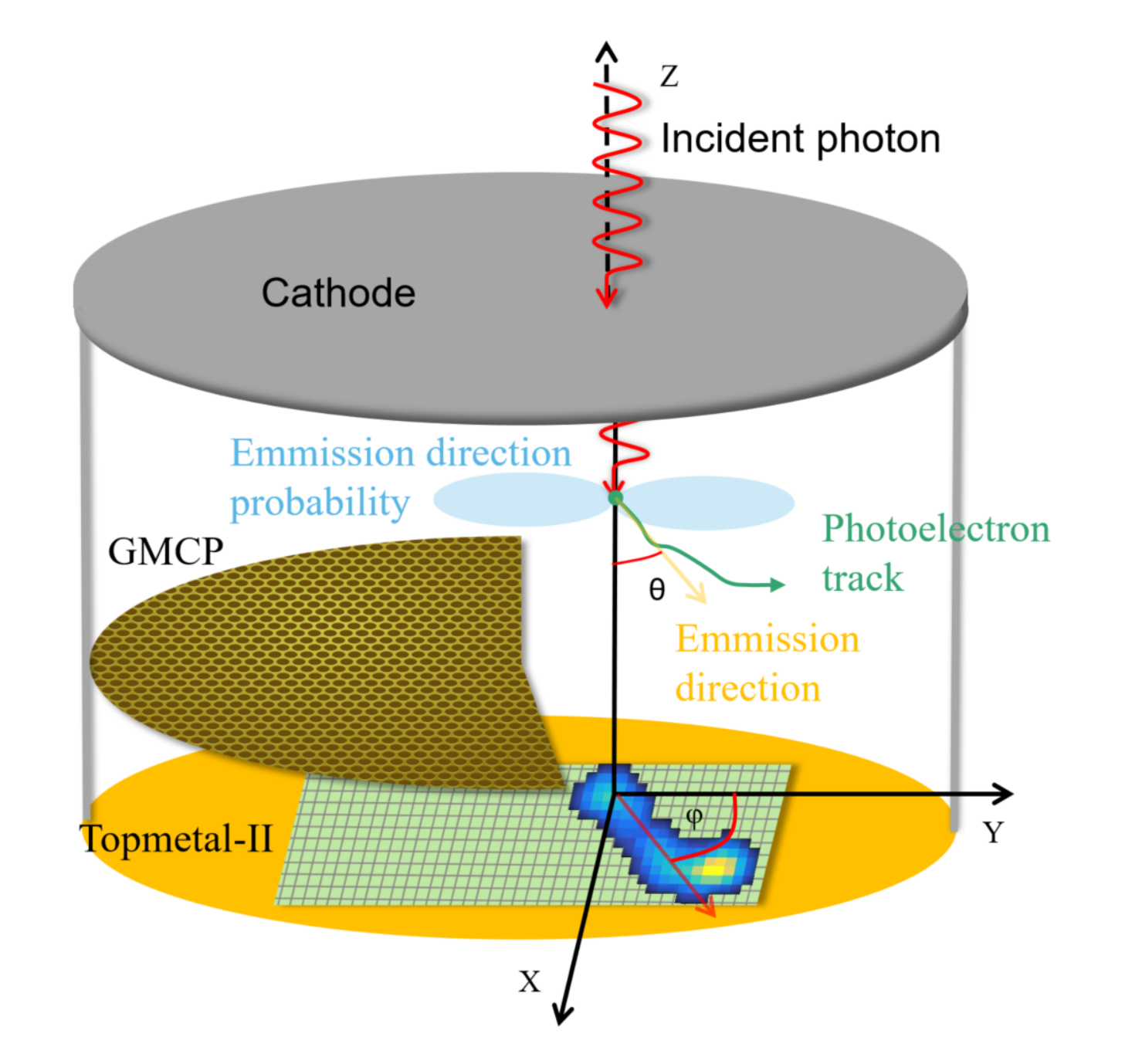}
\caption{Polarization detection principle of GMPD.}
\label{fig:principle}
\end{figure}

As shown in Figure \ref{fig:principle}, soft X-rays pass through the cathode window of the detector unit, and there is a certain probability of photoelectric effect occurring in the drift region, generating photoelectrons carrying the polarization information of the incident photons. These photoelectrons deposit ionization energy in the gas, producing secondary ionization electrons, until they come to a complete stop. In the induction region, an upward electric field is applied, causing some of the secondary ionization electrons to drift downward towards the surface of the GMCP. Some of these electrons enter the microchannels and undergo avalanche multiplication. The multiplied electrons then appear from the bottom surface of the GMCP, with some being absorbed. The remaining multiplied electrons continue to drift towards the Topmetal chip, inducing signals at the corresponding pixel positions. This process projects the track of the photoelectrons onto the two-dimensional plane of the Topmetal chip.

\begin{figure}[htbp]
  \centering
  \subfigure[]{\includegraphics[width=0.35\textwidth]{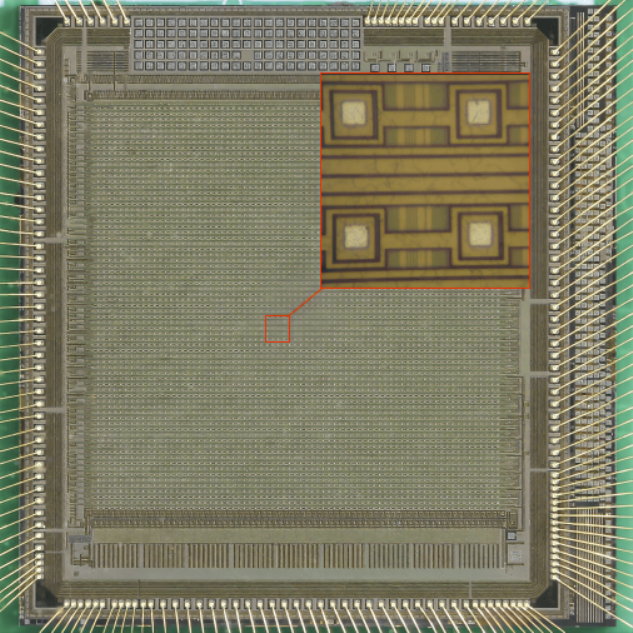}}
  \hfill
  \subfigure[]{\includegraphics[width=0.5\textwidth]{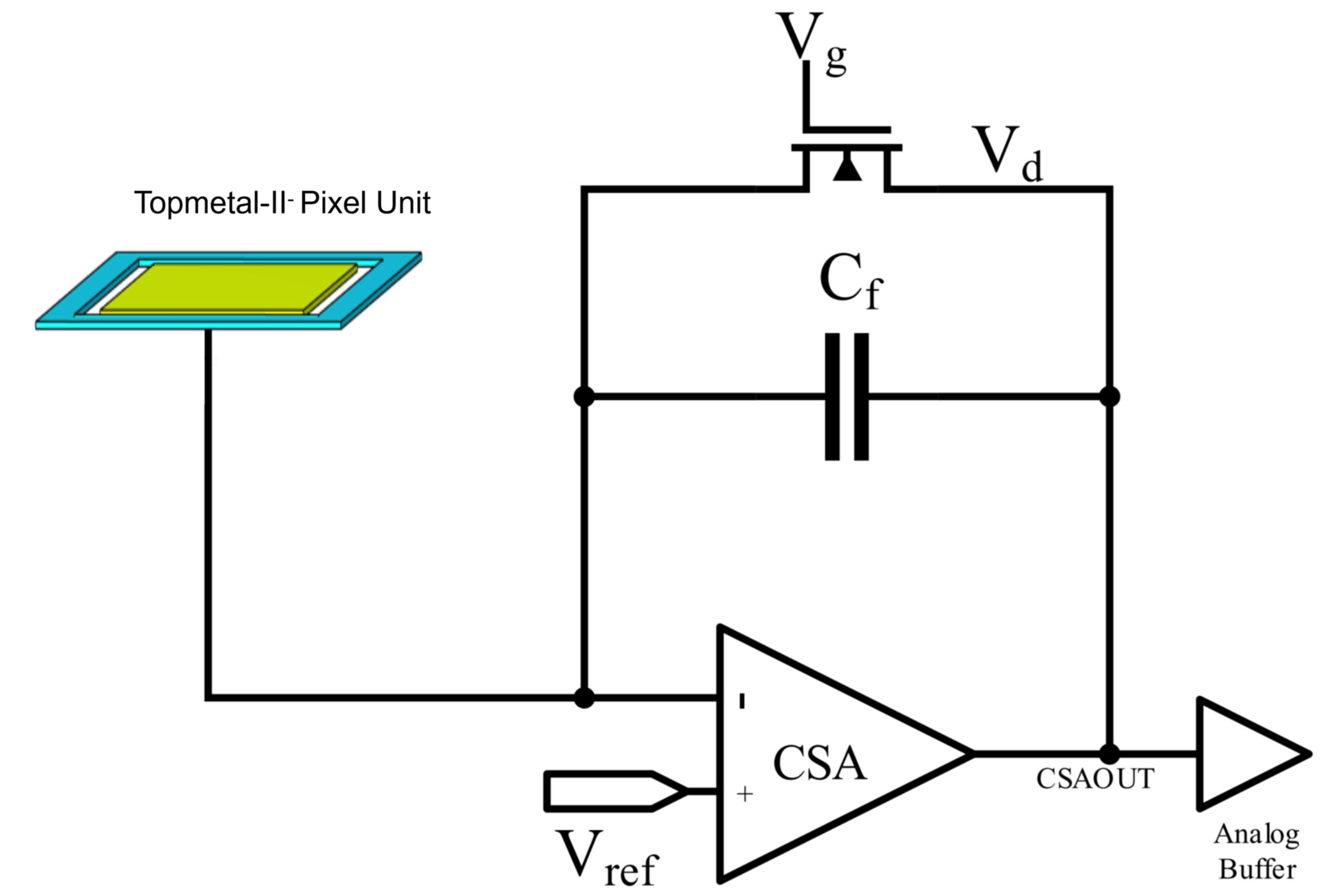}}
  \caption{(a) Top view of the Topmetal-II$^-$. (b) Topmetal-II$^-$ pixel CSA structure. The green portion represents the top metal, which is used to induce the drift charge signal. The blue portion is the Guardring. C$_\text{f}$ denotes the feedback capacitor. V$_\text{g}$ represents the gate voltage, V$_\text{d}$ represents the drain voltage, and V$_\text{ref}$ represents the amplifier's reference voltage.}
  \label{fig:Topmetal-II$^-$}
\end{figure}

The Topmetal-II$^-$ is a pixel sensor manufactured using standard 0.35\,$\mu$m complementary metal-oxide-semiconductor (CMOS) technology. The pixel array is designed in a $72\times72$ pixel matrix pattern with a pixel pitch of 83\,$\mu$m, and features a $6\times6$\,mm charge-sensitive area. The sensor measures $8\times9$\,mm in size, with each pixel's top metal measuring $25\times25$\,$\mu$m. The area of the exposed top metal electrode designed for direct charge induction is $15\times15$\,$\mu$m. Figure \ref{fig:Topmetal-II$^-$}(a) shows a photograph of the sensor bonded to a printed circuit board (PCB) base through wire bonding, as well as an enlarged view of the pixel matrix. Figure \ref{fig:Topmetal-II$^-$}(b) shows the pixel structure of the Topmetal-II$^-$. Each top metal electrode is surrounded by a Guardring, all of which are connected together for performance testing. Gaps and coupling capacitance between the top metal electrode and the Guardring are 3.5\,$\mu$m and 5.5\,fF, respectively. Each exposed electrode is directly connected to a low-noise charge-sensitive amplifier (CSA). When the pixel collects charge, the CSA immediately converts it into an analog output signal, which can then be output through analog or digital readout channels. Analog and testing indicate a charge conversion gain of approximately 196\,mV/fC\cite{Gao_2016}.

\section{Charging-up effect and simulation}
\label{sec:simulation}
\begin{figure*}[htbp]
  \centering
  \subfigure[]{\includegraphics[width=0.29\textwidth]{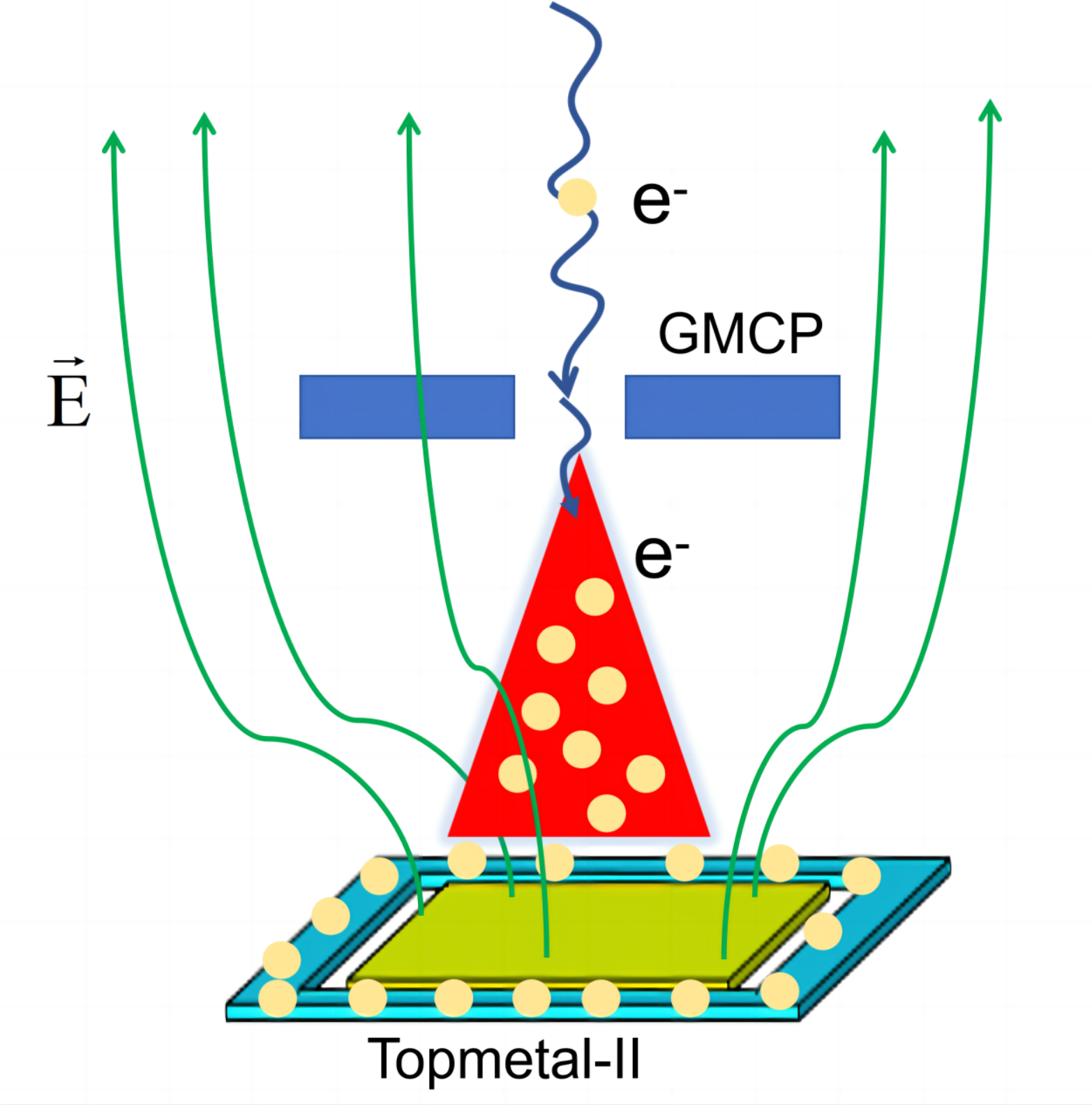}}
  \hfill
  \subfigure[]{\includegraphics[width=0.29\textwidth]{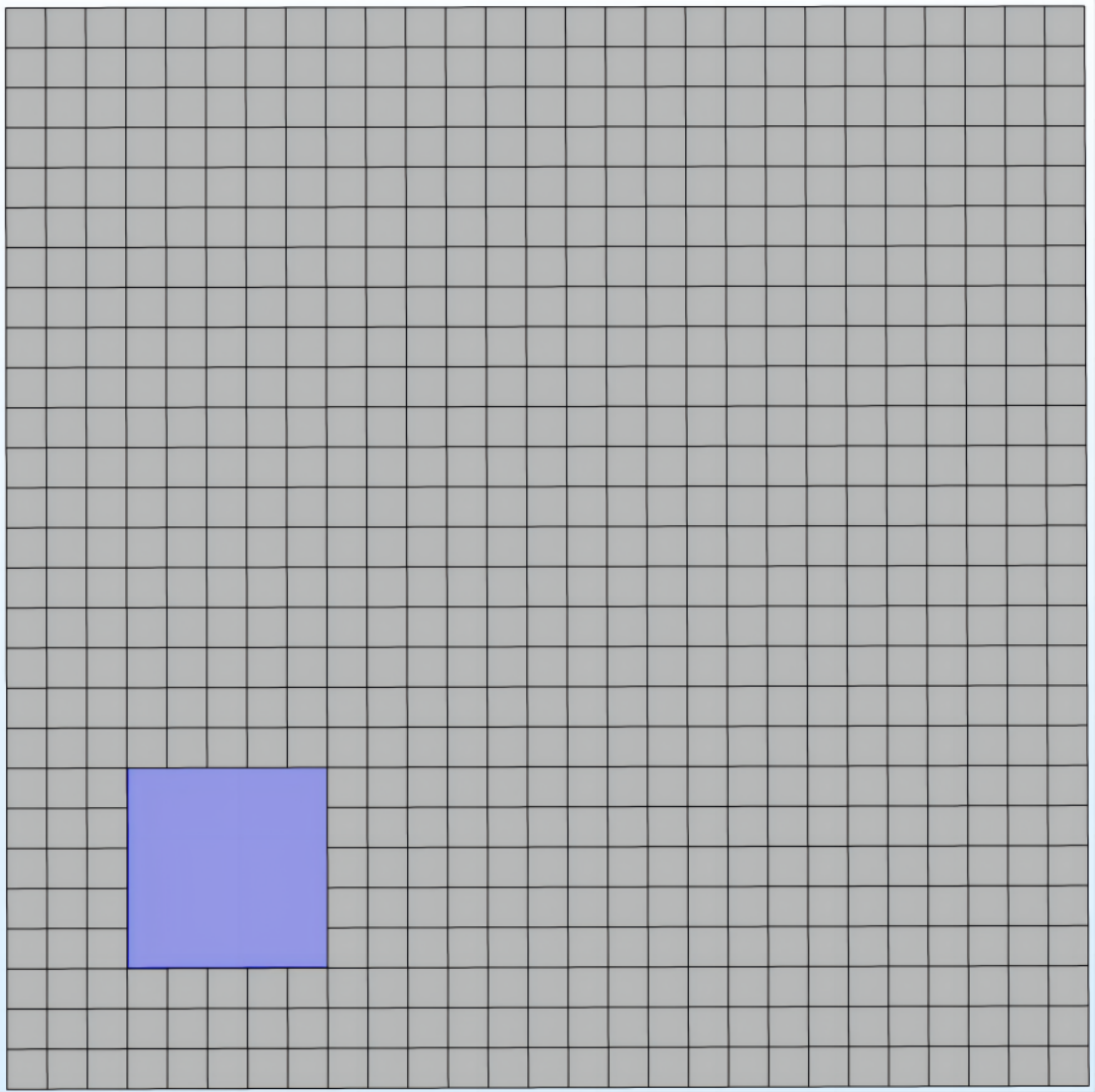}}
  \hfill
  \subfigure[]{\includegraphics[width=0.29\textwidth]{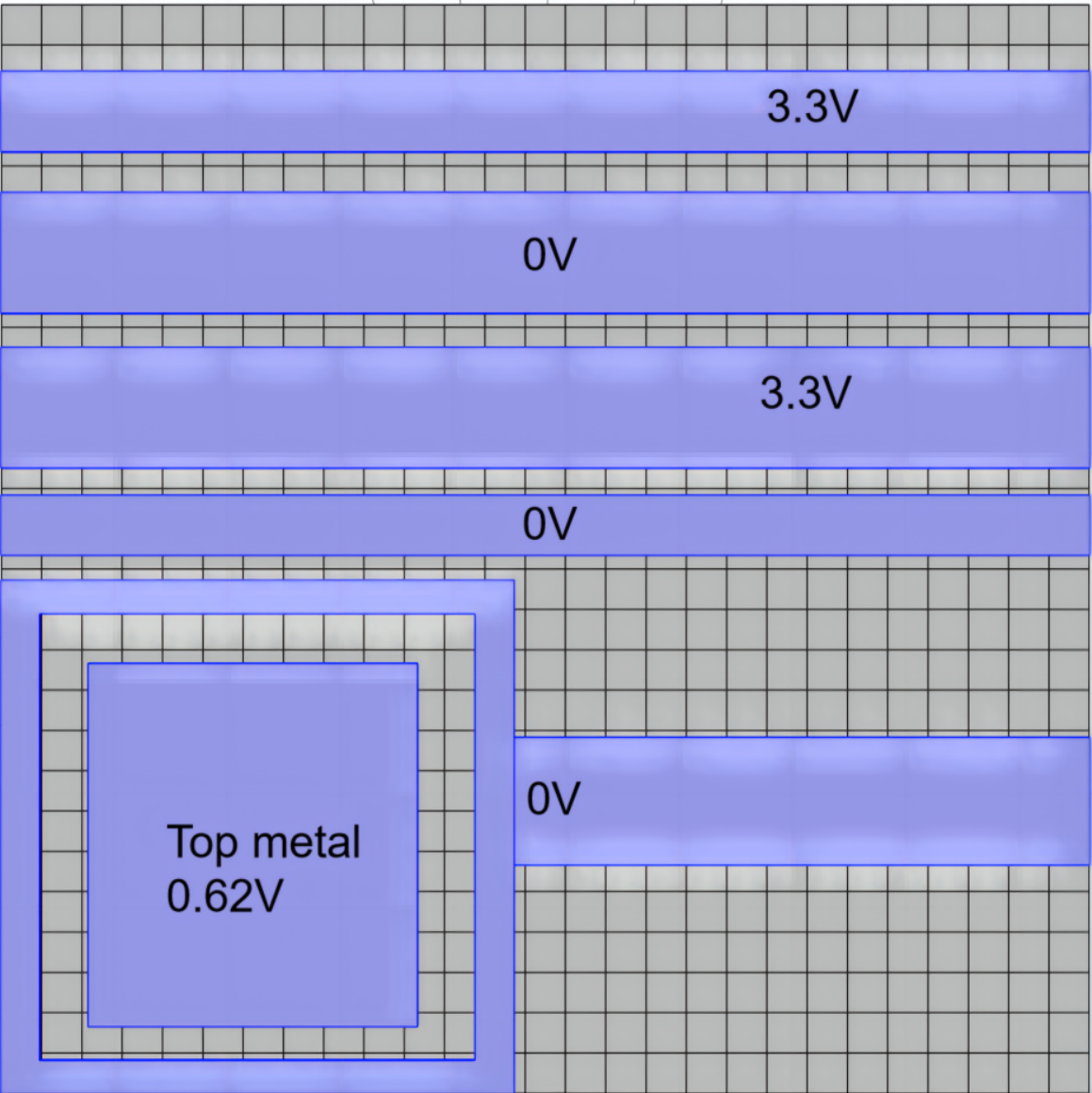}}
  \caption{(a) Charge Accumulation Schematic: Electrons accumulate on the insulating layer, forming a low potential region around the top metal. This leads to the formation of a funnel-shaped electric field above the top metal, thereby enhancing the charge collection efficiency of the pixel. (b) Front view structure of a single pixel, where grey represents the insulating layer where charge accumulates, and blue represents the metal. (c) Back view structure of a single pixel.}
  \label{fig:Accum}
\end{figure*}

The mechanism behind the charging-up effect in Topmetal-II$^-$ is the accumulation of electrons on the surface of the charge induction chip. The surface of the Topmetal-II$^-$ chip has a grid-like insulating layer, which only has $15\times15$\,$\mu$m windows above the top metal. If the charge falls in the window area, it will be conducted away by the top metal, while electrons that fall outside the window area on the insulating layer can hardly move. Therefore, as the number of accumulated events increases during the operation of the detector, the number of charges falling on the insulating layer outside the window area gradually increases, forming an increasingly enhanced electric field in the reverse direction of the induction area as shown in Figure \ref{fig:Accum}(a). At the same time, since the potential of the window area remains unchanged as the potential of the top metal, the electric field lines above this area gradually change from parallel to funnel-shaped, as shown in the figure. The deformed electric field will attract more electrons from outside the window area towards the top metal, ultimately increasing the charge collection efficiency, leading to an effective gain increase in the detector.

To quantitatively understand the impact and variation of charging-up on collection efficiency, we conducted a combined simulation using COMSOL and GARFIELD++ to simulate the change in collection efficiency during the process of charging-up. To simulate the process of charging-up, we followed the following iterative logic:
\begin{itemize}
\item Use COMSOL to model the electric field and potential distribution within the pixel array and detector. As shown in Figure \ref{fig:Accum}(b),(c), to simulate the accumulation of charge at different positions on the pixel surface, we divided the insulating layer surface of a single pixel into 3×3\,$\mu$m patches, initialized the charge on each patch, and calculated and solved for the electric field distribution near the pixel based on the imported surface charge and potential distribution.

\item Import the electric field and geometric model obtained from COMSOL into GARFIELD++. In each iteration, 500 randomly generated electron positions were created at the height below the GMCP, and the process of these electrons drifting in the electric field was simulated in GARFIELD++. The positions where these electrons drifted onto the pixel surface were recorded to simulate the drift of charge and its distribution on the insulating layer of the pixel.

\item Based on the distribution of drifting charge on the pixel, update the accumulated charge on each element. The surface charge density on each element was updated one by one, and the updated results were input into COMSOL to solve for the updated electric field distribution above the pixel array, and steps 2 and 3 are repeated.
\end{itemize}

An automated script was developed to facilitate the iterative process of conversion and updating between COMSOL and GARFIELD++. Following each iteration, we were able to calculate the charge collection efficiency of the pixel under the corresponding charging-up state based on the distribution of the final positions of the electrons.

\section{Modeling and Comparison}
\label{sec:modeling}
We compared the simulation results of the variation in collection efficiency $G(n)$ with experimental measurements. The X-ray source used in the experiment is a 4.51\,keV monochromatic source. For each collected set of photoelectron events, the peak ADC value is obtained by fitting the SumADC distribution induced on the Topmetal-II$^-$ using a crystal ball function. Due to the good linear relationship between the induced voltage on the chip and the output ADC value, the measured variation in peak ADC for the monochromatic source can characterize the change in charge collection efficiency of the pixel.

\begin{figure}[htbp]
\includegraphics
  [width=1.\hsize]
  {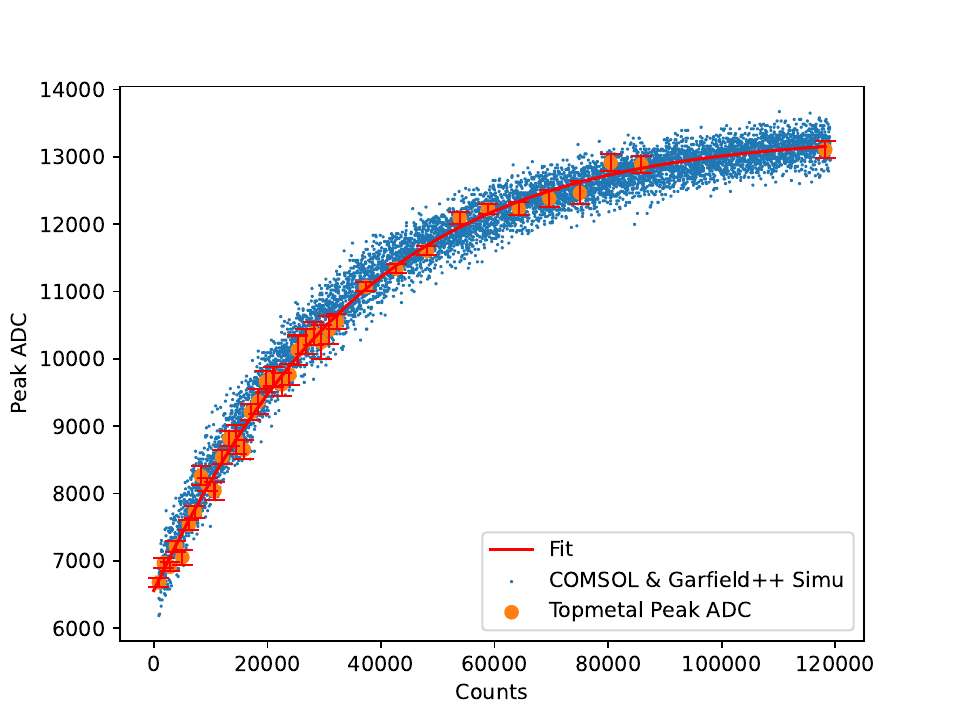}
\caption{The variation of chip collection efficiency with the accumulation of events. The orange points represent experimental data results of 4.51\,keV photoelectron track energy deposition, characterized by fitting the peak ADC of the data spectrum to represent the variation in chip charge collection efficiency. The blue points represent the trend of collection efficiency variation obtained from the joint simulation using COMSOL and GARFIELD++.}
\label{fig:COMSOL_GARFIELD}
\end{figure}

We considered a more simplified charging-up model that describes the relationship between the accumulated event count and the accumulated charge in Equation \ref{eq:charge_Accum}. Where $n$ is the number of events, $q$ is the accumulated charge on the chip, and $q_{\text{max}}$ is the maximum saturated accumulated charge, and $a_c$ is the charge adsorption coefficient. Since the surface of Topmetal-II$^-$ is an insulating layer, the charge dissipation rate is extremely slow, hence we have neglected the charge decay term. The general solution for $q(n)$ is given by equation \ref{eq:qn}.The charge collection efficiency $G(n)$ is proportional to the accumulated charge $q(n)$, but due to the irregular surface structure of the pixel chip and the interconnected circuits with different potentials, the relationship between $G(n)$ and $q(n)$ is difficult to determine analytically. After several attempts, we found that within a certain range of accumulated charge, $G(n)\propto g(n)^2$, as shown by the fitted red line in Figure \ref{fig:COMSOL_GARFIELD}. The experimental and simulated results exhibited good consistency. Through the experiments and simulations, we observed that the change in gain due to charging-up effects gradually saturates as the accumulation events increase. As the accumulated charges in the insulating layer are difficult to dissipate, we found that after placing the detector for a month, the gain remained unchanged within the margin of error.

\begin{equation} \label{eq:charge_Accum} \frac{d q(n)}{d n}=\alpha_c\left(1-\frac{q(n)}{q_{\max }}\right). \end{equation}

\begin{equation} \label{eq:qn} q(n) = x_{0}+x_{1}\exp(x_{2}(n+x_{3})). \end{equation}

\section{Impact on polarization detection}
\label{sec:impact}
\begin{figure}[htbp]
  \centering
  \subfigure[]{\includegraphics[width=0.45\textwidth]{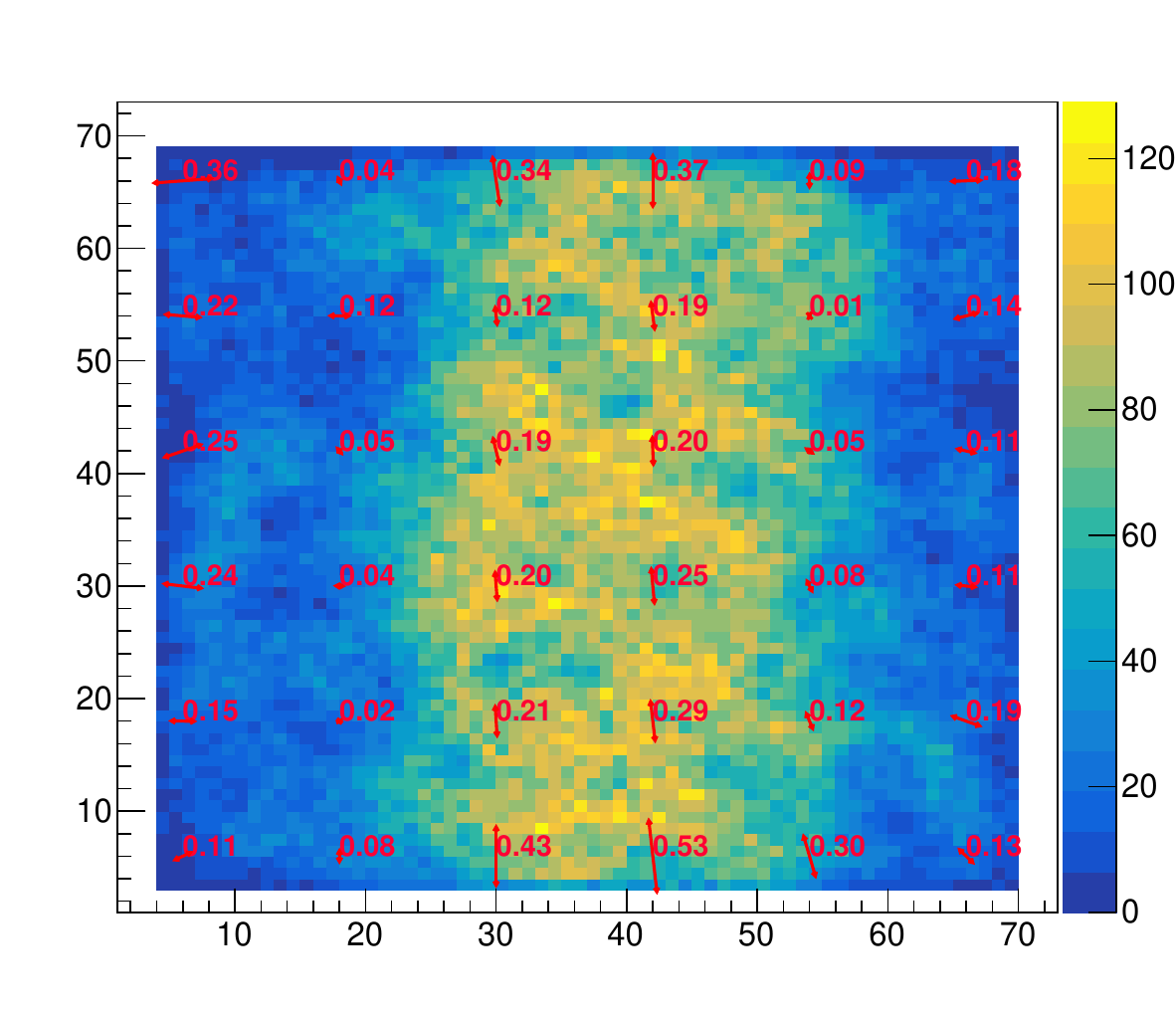}}
  \hfill
  \subfigure[]{\includegraphics[width=0.45\textwidth]{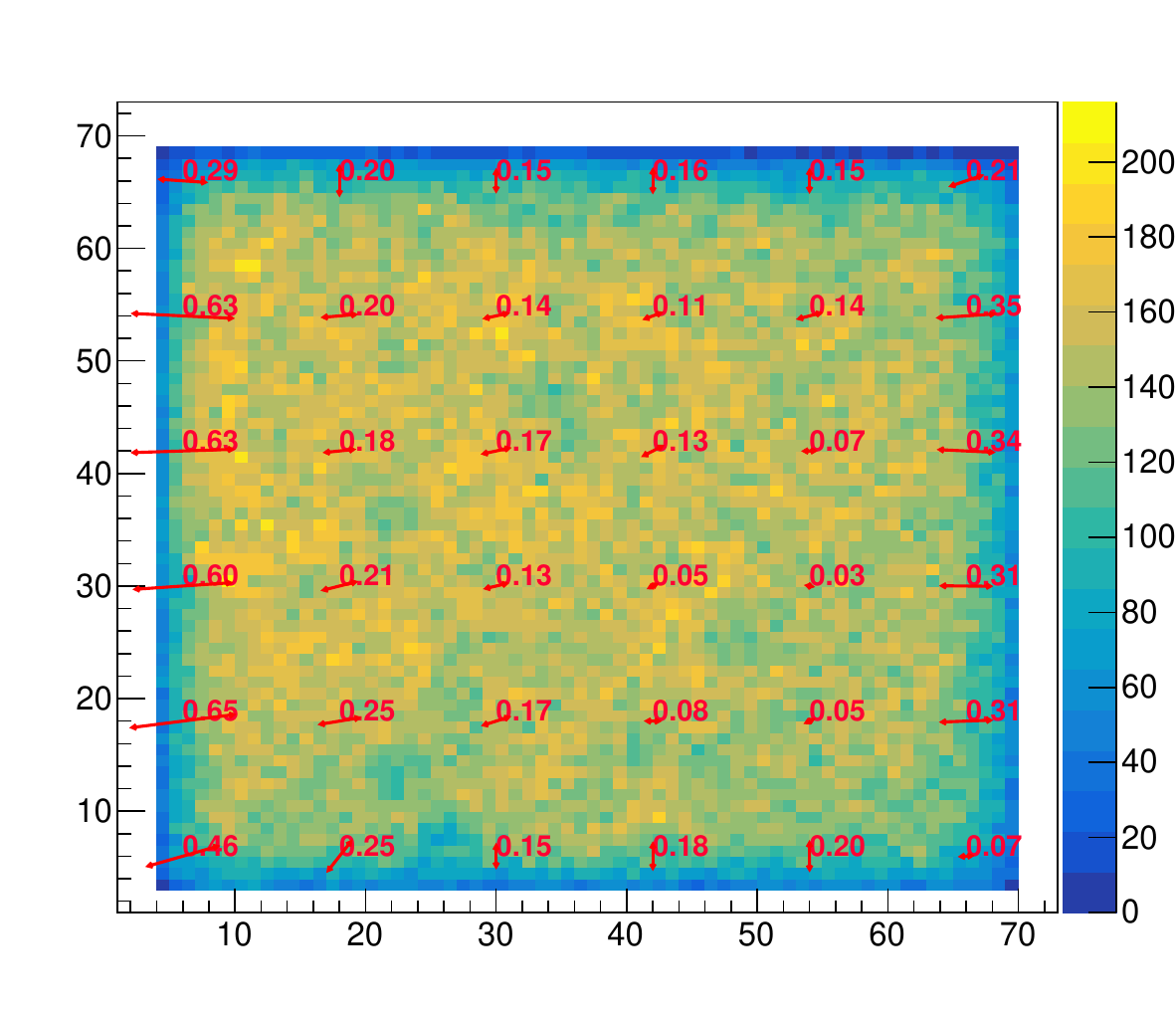}}
  \caption{(a) The residual modulation distribution of 5.9keV Fe$^{55}$ tracks after uneven charging-up due to narrow gap obstruction. (b) The residual modulation distribution of 5.9keV Fe$^{55}$ tracks after uniform charging-up following the removal of the narrow gap obstruction. The heatmap represents the distribution of the reconstructed photoelectron emission positions of the tracks, with the direction of the red lines indicating the direction of residual modulation. The length of the line and the adjacent number represent the value of the residual modulation.}
  \label{fig:Uneven_Accum}
\end{figure}

Charge-up leads to localized gain inconsistencies on the chip, which will affect our reconstruction of the direction of electron tracks, resulting in non-zero residual modulation degree and residual modulation orientation in specific directions. Figure \ref{fig:Uneven_Accum} illustrates the residual modulation caused by the charging-up effect. Initially, a ferrous strip was used to partially obstruct a section of the detector's field of view, leaving a gap of approximately 3\,mm. Following a 2-hour exposure to an X-ray flat source, the obstruction was removed, and a 5.9keV unpolarized Fe$^{55}$ source was used to irradiate and collect the photoelectron tracks. Upon reconstruction, it was observed in Figure \ref{fig:Uneven_Accum}(a) that the signal gain at the previous narrow gap position was significantly higher than the shaded area, and the residual modulation in the narrow gap area was higher than in the shaded area, with the modulation direction parallel to the gap. Subsequently, without any obstruction, the X-ray flat source was used again for 4 hours to accumulate charges on the entire surface of the chip to near saturation. The detector was then irradiated with the 5.9\,keV unpolarized Fe$^{55}$, and the tracks were reconstructed in Figure \ref{fig:Uneven_Accum}(b). Comparing the results of the Fe$^{55}$ measurements before and after charging-up reached saturation, it was found that the residual modulation caused by the uneven gain due to charging-up significantly decreased. Therefore, it is possible to mitigate the impact of the charging-up effect by calibrating or measuring the detector after saturating the charging-up before conducting experiments. Since the accumulated charge is unlikely to naturally dissipate, once the detector is encapsulated, only one thorough charging-up is required.

\section{Suppressing the Charging-up Effect}
\label{sec:suppressing}
Based on the results obtained from the simulations, experiments, and models, the charge accumulation effect can increase the induction strength of charge on pixel pairs by more than twice. This has a significant positive impact on energy measurement and the improvement of track signal-to-noise ratio. However, the charge accumulation effect has challenging implications for X-ray polarimetric detection: on one hand, non-uniform charge accumulation can lead to uneven gain distribution across the chip surface. This can particularly cause significant residual modulation for focusing imaging X-ray polarimeters, as the X-ray source imaging is concentrated in a small area of the chip. On the other hand, with the accumulation of events or the occurrence of discharge processes, the charge accumulation state on the chip surface can change, altering the detector's original energy, position, and polarimetric measurement performance, posing significant calibration difficulties for detectors in orbit.

\begin{figure}[htbp]
\includegraphics
  [width=1.\hsize]
  {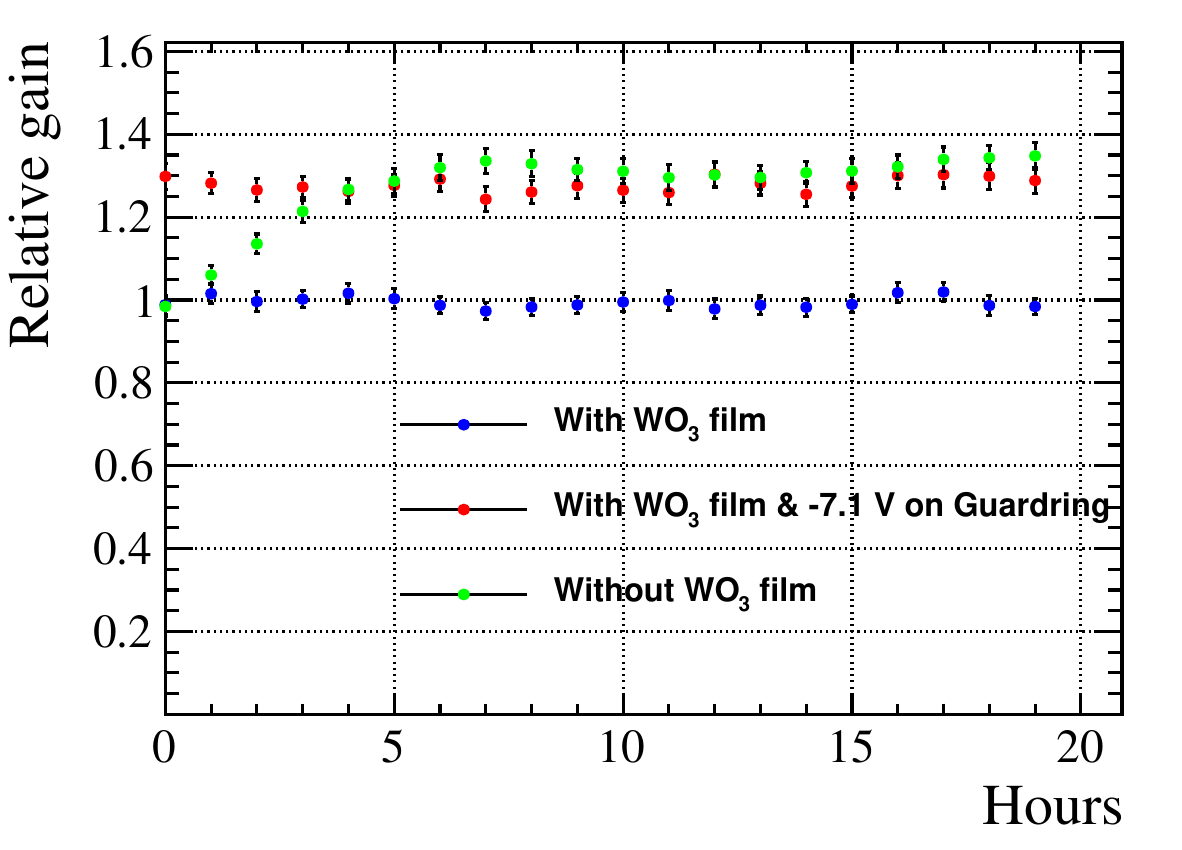}
\caption{Changes in relative gain before and after filming with respect to detector operation time.}
\label{fig:relative_gain}
\end{figure}

Therefore, we test the deposition of a resistive material\cite{FENG2022166595} on the chip surface to release charge and eliminate charge accumulation effects. We use a PVD-75 machine to deposit a layer of WO$_{3}$ resistive material\cite{articleZhen,Bonardo_2021} on the chip surface, with a WO$_{3}$ film thickness of approximately 300\,nm and a resistance of about $10^3$\,M$\Omega$/$\Box$. We conducted gain tests on the chips before and after filming with the WO$_{3}$ layer using a Fe$^{55}$ source for several hours. As a control, a constant voltage of -7.1\,V was applied to the Guardring of the filmed chip to simulate a field region formed by a certain charge accumulation on the chip. The test results, as shown in Figure \ref{fig:relative_gain}, indicate that the gain variation of the chip filmed with the WO$_{3}$ film was less than 5\% during the experiment and did not exhibit a noticeable increasing trend compared to before filming. In the case of applying 7.1 volts, the gain increased by approximately 30\% and remained relatively stable over time. This suggests that when there is a certain demand for gain stability and uniformity, we can eliminate charge accumulation effects on the chip by depositing a resistive layer. For situations where the gain is low without charge accumulation, we can increase the gain and ensure stability by applying a negative bias to the Guardring.

\section{Conclusion}
\label{sec:conclusion}
This study investigates the charging-up effect on the Topmetal-II$^-$ chip in GMPD detectors. We found that this effect differs from the charging-up typically observed in gas detector multiplier devices, and it leads to an increase in the effective gain of the detector. Our research indicates that this effect originates from the accumulation of charges on the insulating layer of the chip's pixel surface. We simulated the process of charging-up on the Topmetal-II$^-$ chip surface and the evolution of pixel charge collection efficiency during the charging-up process. The simulated results align well with the gain variation curve obtained from experimental measurements. By combining experimental data and physical mechanisms, we have proposed a simple model to describe the charging-up effect and the variation in detector gain. This model allows us to calculate the gain under different charging-up states and estimate the magnitude of the saturated gain.

The charging-up effect of Topmetal-II$^-$ can enhance the effective gain of the detector. We also observed that the charges accumulated on the chip's surface are unlikely to dissipate naturally in a short period of time and require the use of metal materials in contact with the chip surface to dissipate the charges, or the use of positive ion wind guns to neutralize the surface charges in order to restore the detector's gain performance to its pre-charging-up state. Therefore, the charging-up effect contributes to the improvement of the detector's performance and stability. However, uneven charging-up effects can cause variations in gain across different regions of the chip, and polarization measurements require the pixel array of the detector to maintain good consistency and stability. Thus, our research results indicate that prior to conducting polarization detection in GMPD, the detector needs to undergo a period of charging-up to ensure stable detector performance and minimize residual modulation in polarization detection. Alternatively, charging-up can be suppressed and relatively high gains can be maintained by filming the chip with resistive materials and adjusting the Guardring voltage.



%%       \citet{<label>} ==> Jones et al. [21]
%%       \citep{<label>} ==> [21]
%%



\bibliographystyle{ieeetr}
\small\bibliography{sample}
\end{document}